\documentclass[12pt]{article}

\usepackage{amsmath,amssymb,cite}
\usepackage{graphicx}

\makeatletter
 
  \@addtoreset{equation}{section}
 \makeatother

\def\XXint#1#2#3{{\setbox0=\hbox{$#1{#2#3}{\int}$} 
\vcenter{\hbox{$#2#3$}}\kern-.5\wd0}}

\allowdisplaybreaks

\begin{document}

\setlength{\oddsidemargin}{0cm}
\setlength{\baselineskip}{7mm}

\begin{titlepage}
\renewcommand{\thefootnote}{\fnsymbol{footnote}}
\begin{normalsize}
\begin{flushright}
\begin{tabular}{l}
KUNS-2255\\
RIKEN-TH-184\\
HRI/ST/1003\\
February 2010
\end{tabular}
\end{flushright}
  \end{normalsize}

~~\\

\vspace*{0cm}
    \begin{Large}
    \begin{bf}
       \begin{center}
         {Large $N$ reduction on coset spaces}
       \end{center}
    \end{bf}   
    \end{Large}
\vspace{0.7cm}

\begin{center}
Hikaru Kawai$^{1),2)}$\footnote
            {
e-mail address : 
hkawai(at)gauge.scphys.kyoto-u.ac.jp}, 
Shinji Shimasaki$^{1),3)}$\footnote
            {
e-mail address : 
shinji(at)gauge.scphys.kyoto-u.ac.jp}
    and
Asato Tsuchiya$^{4)}$\footnote
           {
e-mail address : 
satsuch(at)ipc.shizuoka.ac.jp}\\

\vspace{0.7cm}
                    
 $^{1)}$ {\it Department of Physics, Kyoto University, Kyoto 606-8502, Japan}\\
 \vspace{0.3cm}
 $^{2)}$ {\it Theoretical Physics Laboratory, RIKEN, Wako 351-0198, Japan}\\
 \vspace{0.3cm}
 $^{3)}$ {\it Harish-Chandra Research Institute}\\
 {\it Chhatnag Road, Jhusi, Allahabad 211019, India }\\
\vspace{0.3cm}
 $^{4)}$ {\it Department of Physics, Shizuoka University}\\
 {\it 836 Ohya, Suruga-ku, Shizuoka 422-8529, Japan}
               
\end{center}

\vspace{0.7cm}

\begin{abstract}
\noindent
As an extension of our previous work concerning the large $N$ reduction on group manifolds,
we study the large $N$ reduction on coset spaces. We show that large $N$ field theories on coset spaces
are described by certain corresponding matrix models. We also construct Chern-Simons-like theories on group manifolds
and coset spaces, and give their reduced models. 
\end{abstract}
\vfill
\end{titlepage}
\vfil\eject

\setcounter{footnote}{0}

\section{Introduction}
The large $N$ reduction \cite{EK} asserts that large $N$ field theories are equivalent to certain corresponding matrix models, which are
called the reduced models (for further developments in the large $N$ reduction,
see \cite{Bhanot:1982sh,Parisi:1982gp,Gross:1982at,Das:1982ux,GonzalezArroyo:1982hz,Makeenko:1999hq,
Narayanan:2003fc,Kovtun:2007py,Vairinhos:2007qz,Azeyanagi:2007su,Unsal:2008ch,Bringoltz:2009kb,Poppitz:2009fm}).
In particular, in the case of gauge theories, these matrix models are obtained by dimensionally reducing the original theories
to lower dimensions. The large $N$ reduction is conceptually interesting in the sense that
it realizes emergent space-times.
It is also practically important 
because the reduced models can serve as
a non-perturbative formulation of large $N$ field theories.
However, there is a difficulty. 
Because of  the so-called $U(1)^d$ symmetry breaking \cite{Bhanot:1982sh}, some remedy is needed
in the case of gauge theories. In particular,
no remedy that preserves supersymmetry is known.

While the large $N$ reduction has been studied so far on flat space-times,
it is important to generalize it to curved space-times from both the conceptual and practical viewpoints. 
First, it can provide hints to the problem of describing curved 
space-times \cite{Hanada:2005vr} in the matrix models that are
conjectured to give a non-perturbative formulation of superstring \cite{BFSS,IKKT,DVV}. 
Second, the reduced models on curved space-times are 
in general free from the $U(1)^d$ symmetry breaking.
In particular, 
the reduced models of supersymmetric gauge theories on curved space-times can serve as their non-perturbative
formulation that respects (full) supersymmetry.

Recently, it was shown in \cite{KST} that the large $N$ reduction holds on general group manifolds, which are typical curved manifolds\footnote{
In \cite{Ishii:2008ib}, a different type of the large $N$ reduction on $SU(2)\simeq S^3$ was also
developed. For earlier discussions and further developments, see 
\cite{Ishiki:2006yr,Ishii:2008tm,Ishiki:2008te,Ishiki:2009sg,Kitazawa:2008mx,
Ishiki:2009vr,Ishiki:2010pe,Hanada:2009hd,Hanada:2009kz}.}.
In this paper, we extend it to the case of coset spaces\footnote{While non-commutative field theories on coset spaces such as
$CP^n(\simeq SU(n+1)/SU(n)\times U(1))$\cite{Madore:1991bw,Grosse:1995ar,CarowWatamura:1998jn,Iso:2001mg,
CarowWatamura:2004ct,Alexanian:2001qj,Balachandran:2001dd,Kitazawa:2002xj,
Grosse:2004wm,Dolan:2006tx} have been constructed in terms of matrix models, 
our formulation realizes large $N$ field theories on arbitrary coset spaces.}. 
We give a prescription by which the reduced models of large $N$ field theories 
on coset spaces are obtained from the reduced models of the corresponding theories
on group manifolds. We also generalize Chern-Simons (CS) theories on three-dimensional manifolds
to arbitrary group manifolds and coset spaces,
and give the corresponding reduced models.

This paper is organized as follows. In section 2, as a preparation, we summarize some properties of group manifolds and
coset spaces. In section 3, we briefly review the results in \cite{KST}.
In section 4, we study the large $N$ reduction on coset spaces. In section 5, we construct CS-like
theories on group manifolds and coset spaces, and show that the large $N$ reduction also holds.
Section 6 is devoted to
conclusion and discussion.

\section{Group manifolds and coset spaces}
In this section, we describe some properties of group manifolds and coset spaces which are needed in our analysis.
See also \cite{Ishii:2008tm}.
Let $G$ be a compact connected Lie group and $H$ be a Lie subgroup of $G$.
We put $\mbox{dim}\:G=D$ and $\mbox{dim}\:H=d$. The dimension of the coset space $G/H$ is given by $D-d$.
We use the following indices:
$A,B,\cdots$ run from 1 to $D$, $\alpha,\beta,\cdots$ form $1$ to $D-d$, and
$a,b,\cdots$ from $D-d+1$ to $D$.
$M,N,\cdots$ run from 1 to $D$, $\mu,\nu,\cdots$ from 1 to $D-d$, and $m,n,\cdots$ from $D-d+1$ to $D$.

We take a basis of the Lie algebra of $G$, $t_A$, such
that $t_a$ are a basis of the Lie algebra of $H$.
$t_A$ obey a commutation relation 
\begin{align}
[t_A,t_B]=if_{ABC}t_C, 
\end{align}
where $f_{ABC}$ are 
completely anti-symmetric. 
It follows that $f_{ab\alpha}=0$.
Let $x^M$ be coordinates of the group manifold $G$.
$g(x)\in \:G$ is locally factorized as
\begin{align}
g(x)=h(y)L(\sigma),
\end{align}
where $h(y)\in\:H$, and $y^m$ and $\sigma^{\mu}$ are
coordinates
of $H$ and $G/H$, respectively.
The isometry of $G$ is the $G\times G$ symmetry: one acts on $G$ from the left, while the other from the right.
Only the right $G$ symmetry remains as the isometry of $G/H$.

For $g\in G$, a $D\times D$ matrix $Ad(g)$ is defined by
\begin{align}
g\: t_A\: g^{-1}=t_B\: Ad(g)_{BA}.
\end{align}
$Ad(g)$ is an orthogonal matrix, namely 
\begin{align}
Ad(g)_{AB}Ad(g)_{AC}=\delta_{BC}.
\end{align}
Note that for $h\in H$
\begin{align}
Ad(h)_{\alpha a}=Ad(h)_{a\alpha}=0,
\end{align}
which implies that
\begin{align}
&Ad(h)_{\alpha\beta}Ad(h)_{\alpha\gamma}=\delta_{\beta\gamma}, \nonumber\\
&Ad(h)_{ab}Ad(h)_{ac}=\delta_{bc}.
\label{orthogonal matrix}
\end{align}
$f_{ABC}$ is an invariant third-rank tensor:
\begin{align}
Ad(g)_{AD}Ad(g)_{BE}Ad(g)_{CF}f_{DEF}=f_{ABC}.
\end{align}

We define the right invariant 1-form $E^A_M$ by
\begin{align}
\partial_M g(x)g^{-1}(x)=-iE^A_M(x)\:t_A.
\end{align}
$E^A_M$ satisfy the Maurer-Cartan equation
\begin{align}
\partial_ME^A_N-\partial_NE^A_M-f_{ABC}E^B_ME^C_N=0.
\label{Maurer-Cartan equation}
\end{align}
We also define $e^A_{\mu}$ and $\tilde{e}^a_m$ by
\begin{align}
&\partial_{\mu}L(\sigma)L^{-1}(\sigma)=-ie^A_{\mu}(\sigma)\:t_A,\nonumber\\
&\partial_mh(y)h^{-1}(y)=-i\tilde{e}^a_m(y)\:t_a.
\end{align}
Then, the components of $E^A_M$ are given by
\begin{align}
&E^{\alpha}_{\mu}(x)=Ad(h(y))_{\alpha\beta}e^{\beta}_{\mu}(\sigma), \nonumber\\
&E^a_{\mu}(x)=Ad(h(y))_{ab}e^{b}_{\mu}(\sigma), \nonumber\\
&E^{\alpha}_m(x)=0, \nonumber\\
&E^a_m(x)=\tilde{e}^a_m(y).
\end{align}
$\tilde{e}^a_m(y)$ and $e^{\alpha}_{\mu}(\sigma)$ are viewed as vierbeins of $H$ and $G/H$, respectively, and satisfy
\begin{align}
&\partial_m\tilde{e}^a_n-\partial_n\tilde{e}^a_m-f_{abc}\tilde{e}^b_m\tilde{e}^c_n=0, \nonumber\\
&\partial_{\mu}e^{\alpha}_{\nu}-\partial_{\nu}e^{\alpha}_{\mu}-f_{\alpha AB}e^A_{\mu}e^B_{\nu}=0.
\label{relations for e}
\end{align}
Some algebra gives
\begin{align}
&\frac{\partial}{\partial y^m}Ad(h)_{ab}=\tilde{e}^c_mf_{acd}Ad(h)_{db}, \nonumber\\
&\frac{\partial}{\partial y^m}Ad(h)_{\alpha\beta}=\tilde{e}^a_mf_{\alpha a\gamma}Ad(h)_{\gamma\beta}.
\label{Ad(h)}
\end{align}
A right and left invariant metric of $G$ is defined by
\begin{align}
G_{MN}=E^A_ME^A_N.
\end{align}
It is decomposed as
\begin{align}
ds_G^2=G_{MN}dx^Mdx^N=K_{\mu\nu}d\sigma^{\mu}d\sigma^{\nu}
+(Ad(h)_{ba}\tilde{e}^b_m dy^m+e^a_{\mu}d\sigma^{\mu})^2,
\end{align}
where a right invariant metric of $G/H$, $K_{\mu\nu}$, is given by
\begin{align}
K_{\mu\nu}=e^{\alpha}_{\mu}e^{\alpha}_{\nu}.
\label{k_munu}
\end{align}
When $G$ is viewed as a principal $H$-bundle over $G/H$, $e^a_{\mu}$ correspond to the connection.
The Haar measure of $G$ is defined by
\begin{align}
dg=d^Dx \sqrt{G(x)},
\label{Haar measure}
\end{align}
which is factorized as
\begin{align}
dg=d^{D-d}\sigma d^dy \sqrt{K(\sigma)} \det\tilde{e}^a_m(y).
\label{factrization of Haar measure}
\end{align}

The right invariant Killing vector ${\cal L}_A$ is defined by
\begin{align}
{\cal L}_A=-iE^M_A\frac{\partial}{\partial x^M},
\label{right invariant Killing vector}
\end{align}
where $E^M_A$ are the inverse of $E^A_M$. It generates the left translation, and is expressed as 
\begin{align}
&{\cal L}_a=-i\tilde{e}^m_a\frac{\partial}{\partial y^m}, \nonumber\\
&{\cal L}_{\alpha}=-iAd(h)_{\alpha\beta}e^{\mu}_{\beta}\frac{\partial}{\partial \sigma^{\mu}}
+i\tilde{e}^m_be^c_{\nu}e^{\nu}_{\beta}Ad(h)_{bc}Ad(h)_{\alpha\beta}\frac{\partial}{\partial y^m},
\label{L_a and L_alpha}
\end{align}
where $\tilde{e}^m_a$ and $e^{\mu}_{\beta}$ are the inverses of $\tilde{e}^a_m$ and $e^{\beta}_{\mu}$, respectively.
We denote the Lie derivative along the Killing vector ${\cal L}_A$ by $\delta_A$.
For instance, from (\ref{Maurer-Cartan equation}), we see that
\begin{align}
&\delta_AE^B_M=E^N_A\partial_NE^B_M+\partial_ME^N_AE^B_N=-f_{ABC}E^C_M, \label{Lie derivative of E^A_M}\\
&\delta_AG_{MN}=\delta_AE^B_ME^B_N+E^B_M\delta_AE^B_N=0. 
\label{Lie derivative of G_MN}
\end{align}
The second equation indicates the left invariance of the metric $G_{MN}$.

The spin connection on $G$, $\Omega_M^{AB}$,
is determined by the equation
\begin{align}
\partial_ME_N^A-\partial_NE_M^A+\Omega_M^{AB}E_N^B-\Omega_N^{AB}E_M^B=0.
\end{align}
Comparing this equation with (\ref{Maurer-Cartan equation}), we find that
\begin{align}
\Omega_M^{AB}=\frac{1}{2}f_{ABC}E^C_M.
\label{Omega}
\end{align}
It follows from (\ref{Lie derivative of E^A_M}) that
\begin{align}
\delta_A\Omega_M^{BC}=-f_{ABD}\Omega_M^{DC}-f_{ACD}\Omega_M^{BD}.
\label{Lie derivative of Omega}
\end{align}
(\ref{Lie derivative of E^A_M}) and (\ref{Lie derivative of Omega}) show that 
the Lie derivative accompanied by the local Lorentz transformation keeps $E^A_M$ and $\Omega_M^{AB}$ invariant.
Similarly, the spin connection on $G/H$, $\omega_{\mu}^{\alpha\beta}$, is determined by the equation 
\begin{align}
\partial_{\mu}e_{\nu}^{\alpha}-\partial_{\nu}e_{\mu}^{\alpha}
+\omega_{\mu}^{\alpha\beta}e_{\nu}^{\beta}-\omega_{\nu}^{\alpha\beta}e_{\mu}^{\beta}=0.
\end{align}
$\mbox{}$From (\ref{relations for e}), we find that
\begin{align}
\omega_{\mu}^{\alpha\beta}=\frac{1}{2}f_{\alpha\beta\gamma}e^{\gamma}_{\mu}+f_{\alpha\beta a}e^a_{\mu}.
\end{align}

\section{Large $N$ reduction on group manifolds}
In this section, we briefly review the results in \cite{KST}. 
The statement of the large $N$ reduction on $G$ we showed in \cite{KST} is as follows.
Let a large $N$ matrix field theory be defined on $G$. Its action is given by integration of
a Lagrangian density over $G$ with the Haar measure (\ref{Haar measure}). 
We assume that the theory possesses the right $G$ symmetry. In other words, 
the Lagrangian has no explicit 
dependence on the coordinates of $x^M$ of $G$ if all the derivatives are expressed in terms of ${\cal L}_A$
(\ref{right invariant Killing vector}).  
Then, the planar limit of the theory is described by
the reduced matrix model that is obtained by dropping the coordinate dependence of the fields and replacing 
${\cal L}_A$ by the commutator with the matrix $\hat{L}_A$ given explicitly below.
We emphasize here that the left $G$ symmetry is not necessary for the large $N$ reduction to hold.
As we will see in the next section, this fact is crucial in generalizing the large $N$ reduction to 
the case of coset space $G/H$.

In what follows, we illustrate the large $N$ reduction on group manifolds by considering
$U(N)$ Yang-Mills (YM) theory on $G$
with a real scalar and a Dirac fermion in the adjoint representation.
The action is given by\footnote{We can consider other terms such as higher derivative terms 
and the Yukawa interaction term.}
\begin{align}
&S=S_{YM}+S_s+S_f, \label{S}\\
&S_{YM}=\frac{1}{4\kappa^2}\int d^Dx\sqrt{G}\: G^{MP}G^{NQ}\mbox{Tr}(F_{MN}F_{PQ}), \label{S_YM}\\
&S_s=\frac{1}{\kappa^2}\int d^Dx\sqrt{G}\:\left
(\frac{1}{2}G^{MN}(\partial_M\phi+i[A_M,\phi])(\partial_N\phi+i[A_N,\phi])
+\frac{1}{2}m_s^2\phi^2+\frac{1}{4}\phi^4\right), \label{S_s}\\
&S_f=-\frac{1}{\kappa^2}\int d^Dx\sqrt{G}\:\left(\bar{\psi}\gamma^AE^M_A
\left(\partial_M\psi+i[A_M,\psi]+\frac{1}{4}\Omega_M^{BC}\gamma_{BC}\psi\right)+m_f\bar{\psi}\psi\right),
\label{S_f}
\end{align}
where $A_M$, $\phi$ and $\psi$ are $N\times N$ matrix fields, and $F_{MN}=\partial_MA_N-\partial_NA_M+i[A_M,A_N]$.
By expanding $A_M$ as 
\begin{align}
A_M=E^A_MX_A
\end{align}
and using the equations described in the previous section,
we rewrite (\ref{S_YM}) as
\begin{align}
S_{YM}=-\frac{1}{4\kappa^2}\int dg \: \mbox{Tr}({\cal L}_AX_B-{\cal L}_BX_A-if_{ABC}X_C+[X_A,X_B])^2.
\label{S_YM 2}
\end{align}
In a similar manner, (\ref{S_s}) and (\ref{S_f}) are rewritten as
\begin{align}
&S_s=\frac{1}{\kappa^2}\int dg \:\mbox{Tr}\left(-\frac{1}{2}({\cal L}_A\phi+[X_A,\phi])^2
+\frac{1}{2}m_s^2\phi^2+\frac{1}{4}\phi^4\right), \label{S_s 2}\\
&S_f=-\frac{1}{\kappa^2}\int dg \: \mbox{Tr}
\left(i\bar{\psi}\gamma^A({\cal L}_A\psi+[X_A,\psi])
+\frac{1}{8}f_{ABC}\bar{\psi}\gamma^{ABC}\psi+m_f\bar{\psi}\psi\right).
\label{S_f 2}
\end{align}
The theory possesses the $G\times G$ symmetry, while the left $G$ symmetry is not necessary for the
large $N$ reduction.
We take the planar ('t Hooft) limit in which 
\begin{align}
N\rightarrow\infty,\;\;\kappa\rightarrow 0 \;\;\;
\mbox{with} \;\; \kappa^2N=\lambda \;\; \mbox{fixed},
\label{'t Hooft limit}
\end{align}
where $\lambda$ is the 't Hooft coupling.

To obtain the reduced model, we first define an $n$-dimensional vector space $V_n$  by truncating 
the space of the regular representation of $G$ as follows.
We label the irreducible representations of $G$ by $r$, and denote the representation space of 
the representation $r$ by $V^{[r]}$ and its dimension by $d_r$.
We define a set of the irreducible representations, $I_{\Lambda}$, for a positive number $\Lambda$:
\begin{align}
I_{\Lambda}=\{r;C_2(r)<\Lambda^2\},
\end{align}
where $C_2(r)$ is the second-order Casimir of the representation $r$.
Then, $V_n$ is defined by
\begin{align}
V_n=\bigoplus_{r\in I_{\Lambda}}\underbrace{V^{[r]}\oplus\cdots\oplus V^{[r]}}_{d_r}.
\label{V_n}
\end{align}
Note that the dimension of $V_n$ is given by
\begin{align}
n=\sum_{r\in I_{\Lambda}}d_r^2.
\end{align}
Indeed, the space of the regular representation is obtained by taking the $\Lambda\rightarrow\infty$
limit in (\ref{V_n}).
The $\Lambda\rightarrow\infty$ limit corresponds to the $n\rightarrow\infty$ limit, and 
$\Lambda$ plays the role of a ultraviolet cutoff. We next introduce a $k$-dimensional vector space $W_k$
and consider the tensor product space 
\begin{align}
{\cal V}_N=V_n\otimes W_k,
\end{align}
where $N=nk$ is the dimension of ${\cal V}_N$.

The rule to obtain the reduced model is
\begin{align}
&X_A(g)\rightarrow \hat{X}_A,\;\;\; \phi(g)\rightarrow \hat{\phi},\;\;\; \psi(g)\rightarrow \hat{\psi}, \nonumber\\
&{\cal L}_a\rightarrow [\hat{L}_a,\;], \;\;\; \int dg \rightarrow v,
\label{rule}
\end{align}
where $\hat{X}_A$, $\hat{\phi}$, $\hat{\psi}$ and $\hat{L}_A$ are $N\times N$ hermitian matrices that are
linear operators acting on ${\cal V}_N$. $\hat{L}_A$ take the form
\begin{align}
\hat{L}_A=\left(\bigoplus_{r\in I_{\Lambda}} \underbrace{L^{[r]}_A \oplus \cdots \oplus L^{[r]}_A}_{d_r}
\right)\otimes 1_k,
\end{align}
where $L^{[r]}_A$ are the representation matrices of $t_A$ in the representation $r$.
$v$ is given by
\begin{align}
v=V/n,
\end{align}
where $V$ is the volume of $G$:
\begin{align}
V=\int dg.
\end{align}
Applying (\ref{rule}) to (\ref{S_YM 2}), (\ref{S_s 2}) and (\ref{S_f 2}), we obtain the reduced model of (\ref{S}),
\begin{align}
&S_r=S_{YM,r}+S_{s,r}+S_{f,r}, \label{S_r}\\
&S_{YM,r}=-\frac{v}{4\kappa^2}\mbox{Tr}\left([\hat{L}_A,\hat{X}_B]-[\hat{L}_B,\hat{X}_A]-if_{ABC}\hat{X}_C
+[\hat{X}_A,\hat{X}_B]\right)^2, \\
&S_{s,r}=\frac{v}{\kappa^2} \mbox{Tr}\left(-\frac{1}{2}([\hat{L}_A,\hat{\phi}]+[\hat{X}_A,\hat{\phi}])^2
+\frac{1}{2}m_s^2\hat{\phi}^2+\frac{1}{4}\hat{\phi}^4\right), \label{S_sr}\\
&S_{f,r}=-\frac{v}{\kappa^2}\mbox{Tr}
\left(i\hat{\bar{\psi}}\gamma^A ([\hat{L}_A,\hat{\psi}]+[\hat{X}_A,\hat{\psi}])
+\frac{1}{8}f_{ABC}\hat{\bar{\psi}}\gamma^{ABC}\hat{\psi}+m_f\hat{\bar{\psi}}\hat{\psi}\right).
\label{S_fr}
\end{align}
Making a redefinition
\begin{align}
\hat{L}_A+\hat{X}_A\rightarrow \hat{X}_A
\end{align}
leads to 
\begin{align}
&S_r'=S_{YM,r}'+S_{s,r}'+S_{f,r}', \label{S_r 2}\\
&S_{YM,r}'=-\frac{v}{4\kappa^2}\mbox{Tr}\left([\hat{X}_A,\hat{X}_B]-if_{ABC}\hat{X}_C\right)^2, \label{S_YMr 2}\\
&S_{s,r}'=\frac{v}{\kappa^2}\mbox{Tr}\left(-\frac{1}{2}[\hat{X}_A,\hat{\phi}]^2
+\frac{1}{2}m_s^2\hat{\phi}^2+\frac{1}{4}\hat{\phi}^4\right), \label{S_sr 2}\\
&S_{f,r}'=-\frac{v}{\kappa^2}\mbox{Tr}
\left(i\hat{\bar{\psi}}\gamma^A [\hat{X}_A,\hat{\psi}]
+\frac{1}{8}f_{ABC}\hat{\bar{\psi}}\gamma^{ABC}\hat{\psi}+m_f\hat{\bar{\psi}}\hat{\psi}\right).
\label{S_fr 2}
\end{align}
Note that $S_r'$ is identical to
the dimensional reduction of (\ref{S}) to zero dimension. $\hat{X}_A=\hat{L}_A$ is a classical
solution of $S_r'$, around which we expand $S_r'$ to obtain $S_r$.

The statement of the large $N$ reduction is as follows. Here we assume that $G$ is semi-simple.
If we expand (\ref{S_r 2}) around $\hat{X}_A=\hat{L}_A$, the planar limit of 
(\ref{S}) is retrieved in the limit in which
\begin{align}
n\rightarrow\infty, \;\; k\rightarrow\infty, \;\; \kappa\rightarrow 0,\;\;\;\mbox{with}\;\;
\lambda=\kappa^2N=\kappa^2nk \;\;\mbox{fixed}.
\label{limit for reduced model}
\end{align}
For instance, the correspondence for the free energy is given by
\begin{align}
\frac{F}{N^2V}=\frac{F_r}{N^2v},
\label{relation between two theories}
\end{align}
where $F$ and $F_r$ are the free energies of the original theory and the reduced model, respectively.
For the correspondence for the correlation functions, see \cite{KST}.
The reduced model (\ref{S_r 2}) respects the $G\times G$ symmetry and the gauge symmetry of the original theory.
The latter corresponds to the symmetry given by
\begin{align}
\hat{X}'=U\hat{X}U^{-1}
\label{gauge transformation in reduced model}
\end{align}
for an arbitrary $N\times N$ unitary matrix $U$,
where $\hat{X}$ stands for $\hat{X}_A$ or $\hat{\phi}$ or $\hat{\psi}$ or $\hat{\bar{\psi}}$.

If $G$ is not semi-simple, the above statement does not hold as it stands\footnote{There is no problem for matter fields even if $G$ is not
semi-simple. In fact, (\ref{S_sr}) and (\ref{S_fr}) without $\hat{X}_A$
retrieve the planar limit of (\ref{S_s}) and (\ref{S_f}) without the gauge field, respectively.}. 
The zero-dimensional massless modes around the background $\hat{X}_A=\hat{L}_A$ in (\ref{S_r 2})
makes the background unstable. 
To resolve this problem,
we need a remedy such as the quenching \cite{Bhanot:1982sh,Gross:1982at} or the twisting \cite{GonzalezArroyo:1982hz}.

\section{Large $N$ reduction on $G/H$}
\subsection{Theories on $G/H$ obtained by the dimensional reduction of $G\times G$ symmetric theories on $G$}
In this subsection, we study the large $N$ reduction for theories on $G/H$ that are obtained 
by the dimensional reduction of $G\times G$ symmetric theories on $G$. 
For the dimensional reduction of such theories, see also \cite{Ishii:2008tm}.

Here, as an illustration, we examine
the theory (\ref{S}).
As explained in detail below, the dimensional reduction to $G/H$ is achieved by imposing
the constraints
\begin{align}
&{\cal L}_aX_A=if_{aAB}X_B, \label{constraint for A_M 2}\\
&{\cal L}_a\phi=0, \label{constraint for phi 2}\\
&{\cal L}_a\psi=\frac{i}{4}f_{aAB}\gamma^{AB}\psi,\;\;\;{\cal L}_a\bar{\psi}=-\frac{i}{4}f_{aAB}\bar{\psi}\gamma^{AB}
\label{constraint for psi 2}
\end{align}
on the theory. These constraints are, for instance, realized by
adding 
\begin{align}
&\int dg \mbox{Tr}(M_g^2({\cal L}_aX_B-if_{aBC}X_C)^2+M_s^2({\cal L}_a\phi)^2 \nonumber\\
&\qquad+M_f({\cal L}_a\bar{\psi}+\frac{i}{4}f_{aAB}\bar{\psi}\gamma^{AB})
({\cal L}_a\psi-\frac{i}{4}f_{aCD}\gamma^{CD}\psi))
\label{mass terms}
\end{align}
to the action and taking the $M_g,M_s,M_f\rightarrow\infty$ limit.
Because of these constraints, the $G\times G$ symmetry of (\ref{S}) is broken to the right $G$ symmetry.
As emphasized in the beginning of section 3, the right $G$ symmetry is sufficient for
the large $N$ reduction to hold. The large $N$ reduction, therefore, holds for the theory obtained
by dimensionally reducing (\ref{S}) to $G/H$ as follows.
Applying the rule (\ref{rule}) to the theory (\ref{S}) with
(\ref{constraint for A_M 2}), (\ref{constraint for phi 2}) and
(\ref{constraint for psi 2}) leads to imposing constraints
\begin{align}
&[\hat{L}_a,\hat{X}_B]=if_{aBC}\hat{X}_C, \label{constraint on X_A}\\
&[\hat{L}_a,\hat{\phi}]=0, \label{constraint on phi}\\
&[L_a,\psi]=\frac{i}{4}f_{aCD}\gamma^{CD}\psi,\;\;\;
[\hat{L}_a,\hat{\bar{\psi}}]=-\frac{i}{4}f_{aAB}\hat{\bar{\psi}}\gamma^{AB}
\label{constraint on psi}
\end{align}
on (\ref{S_r}) or (\ref{S_r 2}).
Note that the redefinition $\hat{L}_{\alpha}+X_{\alpha}\rightarrow \hat{X}_{\alpha}$ keeps 
the constraint (\ref{constraint on X_A}) invariant.
For instance, these constrains are realized  by adding
\begin{align}
&\mbox{Tr}(M_g^2([\hat{L}_a.\hat{X}_B]-if_{aBC}\hat{X}_C)^2+M_s^2[\hat{L}_a,\hat{\phi}]^2 \nonumber\\
&\qquad+M_f([\hat{L}_a,\hat{\bar{\psi}}]+\frac{i}{4}f_{aAB}\hat{\bar{\psi}}\gamma^{AB})
([L_a,\psi]-\frac{i}{4}f_{aCD}\gamma^{CD}\psi))
\label{mass terms in reduced model}
\end{align}
to (\ref{S_r}) or (\ref{S_r 2}) and taking the $M_g,M_s,M_f\rightarrow\infty$ limit.
$\hat{X}_A=\hat{L}_A$ satisfies (\ref{constraint on X_A}), (\ref{constraint on phi}) and (\ref{constraint on psi})
and is a classical solution of (\ref{S_r 2}) with (\ref{mass terms in reduced model}). We expand
(\ref{S_r 2}) with (\ref{mass terms in reduced model}) around the classical solution to obtain (\ref{S_r}) with
(\ref{mass terms in reduced model}).
To summarize, the reduced model of the theory on $G/H$ is the matrix model  (\ref{S_r 2})
with the constraints (\ref{constraint on X_A}), (\ref{constraint on phi}) and (\ref{constraint on psi}).
The reduced model retrieves the planar limit of the theory on $G/H$ in the limit (\ref{limit for reduced model}).
It respects the right $G$ symmetry of the theory on $G/H$.
It also has the gauge symmetry  (\ref{gauge transformation in reduced model}) with the 
constraint
\begin{align}
[L_a,U]=0
\label{constraint for gauge transformation}
\end{align}
satisfied. This corresponds to the gauge symmetry of the theory on $G/H$.

In what follows, we see that imposing (\ref{constraint for A_M 2}), (\ref{constraint for phi 2}) and
(\ref{constraint for psi 2}) on (\ref{S}) indeed yields the dimensional reduction to $G/H$.
The left $G$ symmetry corresponds to the invariance of (\ref{S}) under the transformation
\begin{align}
&A_M\rightarrow A_M+\epsilon\delta_AA_M, \label{transformation of A_M}\\
&\phi\rightarrow \phi+\epsilon \delta_A\phi, \label{transformation of phi}\\
&\psi\rightarrow \psi+\epsilon\left(\delta_A\psi+\frac{1}{4}f_{ABC}\gamma^{BC}\psi\right), \;\;\;
\bar{\psi}\rightarrow \bar{\psi}+\epsilon\left(\delta_A\bar{\psi}-\frac{1}{4}f_{ABC}\bar{\psi}\gamma^{BC}\right).
\label{transformation of psi}
\end{align}
This invariance follows from (\ref{Lie derivative of E^A_M}),
(\ref{Lie derivative of G_MN}) and (\ref{Lie derivative of Omega}).
Note that the transformation of the fermion includes 
the local Lorentz transformation as well as the Lie derivative, because
$E^A_M$ and $\Omega_M^{AB}$ are invariant under such a transformation. 
By using ${\cal L}_A$, (\ref{transformation of A_M}), (\ref{transformation of phi}) and (\ref{transformation of psi}) are
expressed as 
\begin{align}
&X_B \rightarrow X_B+\epsilon(i{\cal L}_AX_B+f_{ABC}X_C), \label{transformation of X_B}\\
&\phi \rightarrow \phi+\epsilon i {\cal L}_A\phi, \\
&\psi \rightarrow \psi+\epsilon\left(i{\cal L}_A\psi+\frac{1}{4}f_{ABC}\gamma^{BC}\psi\right), \;\;\;
\bar{\psi} \rightarrow \bar{\psi}+\epsilon\left(i{\cal L}_A\bar{\psi}-\frac{1}{4}f_{ABC}\bar{\psi}\gamma^{BC}\right).
\end{align}
Hence, by imposing the constraints (\ref{constraint for A_M 2}), (\ref{constraint for phi 2}) and (\ref{constraint for psi 2})
on (\ref{S}),
we can make a dimensional reduction from $G$ to $G/H$, which is the so-called consistent truncation.
Namely, every solution to the equation of motion in the dimensionally reduced theory is also
a solution to the equation of motion in the original theory. 

Let us obtain the explicit form of the resultant theory on $G/H$.
Using the equations described in section 2, we solve (\ref{constraint for A_M 2}) as
\begin{align}
&X_{\alpha}=Ad(h(y))_{\alpha\beta}e^{\mu}_{\beta}(\sigma)a_{\mu}(\sigma), \label{solution of constraint for X_alpha}\\
&X_a=-Ad(h(y))_{ab}\phi_b(\sigma).
\label{solution of constraint for X_a}
\end{align}
Similarly, (\ref{constraint for phi 2}) is solved as
\begin{align}
\phi=\phi(\sigma).
\label{solution of constraint for phi}
\end{align}
To solve (\ref{constraint for psi 2}), we introduce $\rho(t_A)$ defined by
\begin{align}
\rho(t_A)=-\frac{i}{4}f_{ABC}\gamma^{BC}.
\end{align}
It satisfies 
\begin{align}
[\rho(t_A),\rho(t_B)]=if_{ABC}\rho(t_C).
\end{align}
Then, we can solve (\ref{constraint for psi 2})  as
\begin{align}
\psi=e^{i\theta^a(y)\rho(t_a)}\chi(\sigma), \;\;\;
\bar{\psi}=\bar{\chi}(\sigma)e^{-i\theta^a(y)\rho(t_a)},
\label{solution of constraint for psi}
\end{align}
where $\theta^a(y)$ is defined by
\begin{align}
h=e^{i\theta^a(y)t_a}.
\end{align}
Substituting (\ref{solution of constraint for X_alpha}) 
and (\ref{solution of constraint for X_a}) into (\ref{S_YM}) leads to
\begin{align}
S^{G/H}_{YM}=&\frac{w}{\kappa^2}\int d^{D-d}\sigma\sqrt{K}\:
\mbox{Tr}\left(\frac{1}{4}(f_{abc}\phi_c+i[\phi_a,\phi_b])^2 \right.\nonumber\\
&\qquad+\frac{1}{2}K^{\mu\nu}(\partial_{\mu}\phi_a+i[a_{\mu},\phi_a]-e^b_{\mu}f_{abc}\phi_c)
                             (\partial_{\nu}\phi_a+i[a_{\nu},\phi_a]-e^d_{\nu}f_{ade}\phi_e)  \nonumber\\
&\left.\qquad+\frac{1}{4}K^{\mu\lambda}K^{\nu\rho}
(f_{\mu\nu}-b^a_{\mu\nu}\phi_a)(f_{\lambda\rho}-b^b_{\lambda\rho}\phi_b)\right),
\label{S^G/H_YM}
\end{align}
where $w$ is the volume of $H$, $f_{\mu\nu}=\partial_{\mu}a_{\nu}-\partial_{\nu}a_{\mu}+i[a_{\mu},a_{\nu}]$,
and $b^a_{\mu\nu}=\partial_{\mu}e^a_{\nu}-\partial_{\nu}e^a_{\mu}-f_{abc}e^b_{\mu}e^c_{\nu}
=f_{a\alpha\beta}e^{\alpha}_{\mu}e^{\beta}_{\nu}$.
The final expression is indeed independent of $y$. We have obtained YM theory coupled to $d$ Higgs fields on $G/H$.
This result agrees with the one in \cite{Ishii:2008tm}.
Similarly, substituting (\ref{solution of constraint for X_alpha}), (\ref{solution of constraint for X_a})
and (\ref{solution of constraint for phi}) into (\ref{S_s}), we obtain
\begin{align}
S^{G/H}_s=&\frac{w}{\kappa^2}\int d^{D-d}\sigma \sqrt{K}\:\mbox{Tr}\left(
\frac{1}{2}K^{\mu\nu}(\partial_{\mu}\phi+i[a_{\mu},\phi])(\partial_{\nu}\phi+i[a_{\nu},\phi]) \right. \nonumber\\
&\qquad\qquad\qquad\left.-\frac{1}{2}[\phi_a,\phi]^2+\frac{1}{2}m_s^2\phi^2+\frac{1}{4}\phi^4\right).
\end{align}
Finally, by using (\ref{solution of constraint for X_alpha}), (\ref{solution of constraint for X_a}), 
(\ref{solution of constraint for psi})  and the equation
\begin{align}
e^{-i\theta^a(y)\rho(t_a)}\gamma_Ae^{i\theta^b(y)\rho(t_b)}=Ad(h)_{AB}\gamma_B,
\end{align} 
(\ref{S_f})  becomes
\begin{align}
S^{G/H}_f=&-\frac{w}{\kappa^2}\int d^{D-d}\sigma\sqrt{k}\:\mbox{Tr}
\left(e^{\mu}_{\alpha}\bar{\chi}\gamma^{\alpha}\left(\partial_{\mu}\chi
+\frac{1}{4}\omega^{\beta\gamma}_{\mu}\gamma_{\beta\gamma}\chi+i[a_{\mu},\chi]\right)\right. \nonumber\\
&\qquad\qquad-i\bar{\chi}\gamma^a[\phi_a,\chi]
+\frac{1}{4}f_{abc}e^{\mu}_{\alpha}e^c_{\mu}\bar{\chi}\gamma^{ab\alpha}\chi \nonumber\\
&\qquad\qquad\left. -\frac{1}{8}f_{abc}\bar{\chi}\gamma^{abc}\chi
+\frac{1}{8}f_{a\alpha\beta}\bar{\chi}\gamma^{a\alpha\beta}\chi+m_f\bar{\chi}\chi\right).
\label{S^G/H_f}
\end{align}
We have obtained $2^{\frac{d}{2}}$-flavor fermions for even $d$, $2^{\frac{d+1}{2}}$-flavor fermions for odd $d$ and even $D$, and 
$2^{\frac{d-1}{2}}$-flavor fermions for odd $d$ and odd $D$.

\subsection{Minimal theories on $G/H$}
In the previous subsection, we obtained the reduced model of
(\ref{S^G/H_YM}), which is YM theory with $d$ Higgs scalars on $G/H$ 
originating from the consistent truncation of pure YM theory on $G$.
We also obtained the reduced model of multi-flavor fermions on $G/H$ originating from
one-flavor fermion on $G$.
In this subsection, we will study the large $N$ reduction for ``minimal"theories on $G/H$: pure YM theory on $G/H$ and 
one-flavor fermion on $G/H$.

We first study pure YM theory on $G/H$. 
As explained below, it is equivalent to 
a theory on $G$ 
\begin{align}
S_{YM}^{min}=-\frac{1}{4\kappa^2}\int dg \: \mbox{Tr}({\cal L}_{\alpha}X_{\beta}-{\cal L}_{\beta}X_{\alpha}
-if_{\alpha\beta\gamma}X_{\gamma}+[X_{\alpha},X_{\beta}])^2.
\label{minimal YM}
\end{align}
with the constraint
\begin{align}
{\cal L}_aX_{\alpha}=if_{a\alpha\beta}X_{\beta}.
\label{constraint for X_alpha}
\end{align}
The theory (\ref{minimal YM}) with (\ref{constraint for X_alpha}) possesses the right $G$ symmetry.
Hence, the large $N$ reduction holds for pure YM theory on $G/H$.
Applying the rule (\ref{rule}) to (\ref{minimal YM}) with (\ref{constraint for X_alpha}),
we obtain the reduced model of pure YM on $G/H$ which is a matrix model
\begin{align}
S^{min}_{YM,r}=-\frac{v}{4\kappa^2}\mbox{Tr}([\hat{L}_{\alpha},\hat{X}_{\beta}]-[\hat{L}_{\beta},\hat{X}_{\alpha}]
-if_{\alpha\beta\gamma}\hat{X}_{\gamma})^2
\label{reduced model of pure YM on G/H}
\end{align}
with the constraint
\begin{align}
[\hat{L}_a,\hat{X}_{\alpha}]=if_{a\alpha\beta}\hat{X}_{\beta}.
\label{constraint for pure YM on G/H}
\end{align}
The reduced model
retrieves the planar limit of pure YM theory on $G/H$
in the limit (\ref{limit for reduced model}).
As before, the redefinition $\hat{L}_{\alpha}+\hat{X}_{\alpha}\rightarrow \hat{X}_{\alpha}$
in (\ref{reduced model of pure YM on G/H}) yields 
\begin{align}
S^{min}_{YM,r}{}'=-\frac{v}{4\kappa^2}\mbox{Tr}([\hat{X}_{\alpha},\hat{X}_{\beta}]
-if_{\alpha\beta\gamma}\hat{X}_{\gamma}-if_{\alpha\beta a}\hat{L}_a)^2.
\label{reduced model of pure YM on G/H 2}
\end{align}
Note again that the redefinition keeps the constraint (\ref{constraint for pure YM on G/H}) invariant. 
Hence, the reduced model of pure YM theory on $G/H$ is also given by the matrix model 
(\ref{reduced model of pure YM on G/H 2}) with  the constraint (\ref{constraint for pure YM on G/H}).
$\hat{X}_{\alpha}=\hat{L}_{\alpha}$ satisfies the constraint (\ref{constraint for pure YM on G/H})
and is a classical solution of the reduced model, 
(\ref{reduced model of pure YM on G/H 2}) with (\ref{constraint for pure YM on G/H}).
We expand the reduced model around the classical
solution and take the limit (\ref{limit for reduced model}) 
to obtain the planar limit of pure YM theory on $G/H$.
Note that (\ref{reduced model of pure YM on G/H 2}) with (\ref{constraint for pure YM on G/H}) is obtained
by putting $\hat{X}_a=\hat{L}_a$ in (\ref{S_YMr 2}) with (\ref{constraint on X_A}).
The reduced model, (\ref{reduced model of pure YM on G/H 2}) with 
(\ref{constraint for pure YM on G/H}), respects the right $G$ symmetry and the gauge symmetry of
pure YM theory on $G/H$.

Now let us see that (\ref{minimal YM}) with (\ref{constraint for X_alpha}) indeed yields pure YM theory on $G/H$.
(\ref{minimal YM}) is obtained by putting
\begin{align}
X_a=0 
\label{X_a=0}
\end{align}
and (\ref{constraint for X_alpha})
in (\ref{S_YM 2}). Recall that (\ref{S_YM 2}) is invariant under the transformation (\ref{transformation of X_B})
with $A=a$. 
Note also that (\ref{X_a=0}) and (\ref{constraint for X_alpha}) are invariant under this transformation.
Hence, (\ref{minimal YM}) has the symmetry given by
\begin{align}
X_{\alpha}\rightarrow X_{\alpha}+\epsilon(i{\cal L}_aX_{\alpha}+f_{a\alpha\beta}X_{\beta}).
\label{transformation for minimal YM}
\end{align}
This implies that we can impose the constraint (\ref{constraint for X_alpha}) on (\ref{minimal YM}) to
truncate (\ref{minimal YM}) consistently to a theory on $G/H$. The solution of 
the constraint (\ref{constraint for X_alpha}) is
given in (\ref{solution of constraint for X_alpha}). By substituting the solution into
(\ref{minimal YM}), we indeed obtain pure YM theory on $G/H$
\begin{align}
S^{min}_{YM}=\frac{w}{4\kappa^2}\int d^{D-d}\sigma\sqrt{K}\:K^{\mu\lambda}K^{\nu\rho}
\mbox{Tr}(f_{\mu\nu}f_{\lambda\rho}).
\label{pure YM on G/H}
\end{align}

Next, we study one-flavor fermion on $G/H$. 
Instead of (\ref{S_f 2}), we consider the following theory on $G$
\begin{align}
S_f'=-\frac{1}{\kappa^2}\int dg \: \mbox{Tr}
\left(i\bar{\psi}\gamma^{\alpha}({\cal L}_{\alpha}\psi+[X_{\alpha},\psi])
+\frac{1}{8}f_{\alpha\beta\gamma}\bar{\psi}\gamma^{\alpha\beta\gamma}\psi+m_f\bar{\psi}\psi\right),
\label{S_f'}
\end{align}
with the constraints (\ref{constraint for X_alpha}) and
\begin{align}
{\cal L}_a\psi=\frac{i}{4}f_{a\alpha\beta}\gamma^{\alpha\beta}\psi,\;\;\;
{\cal L}_a\bar{\psi}=-\frac{i}{4}f_{a\alpha\beta}\bar{\psi}\gamma^{\alpha\beta},
\label{constraint for psi 3} 
\end{align}
where $\psi$ and $\bar{\psi}$ are a $2^{\frac{D-d}{2}}$-component fermion for even $D-d$ and
a $2^{\frac{D-d-1}{2}}$-component fermion for odd $D-d$. Indeed, while (\ref{S_f'}) is a theory on $G$,
$\gamma^{\alpha}$ are the gamma matrices in $D-d$ dimensions.
As we will see below, the theory (\ref{S_f'}) with these constrains represents
one-flavor fermion on $G/H$.
It possesses the right $G$ symmetry, so that the large $N$ reduction holds for it
as in the case of pure YM theory on $G/H$.
Applying the rule (\ref{rule}) to (\ref{S_f'}) with (\ref{constraint for X_alpha}) and (\ref{constraint for psi 3}) and
making the redefinition $\hat{L}_{\alpha}+\hat{X}_{\alpha}\rightarrow \hat{X}_{\alpha}$, we obtain the reduced model
of one-flavor fermion on $G/H$ which is a matrix model
\begin{align} 
S^{min}_{f,r}=-\frac{v}{\kappa^2}\mbox{Tr}\left(i\bar{\psi}\gamma^{\alpha}[\hat{X}_{\alpha},\psi]
+\frac{1}{8}f_{\alpha\beta\gamma}\bar{\psi}\gamma^{\alpha\beta\gamma}\psi+m_f\bar{\psi}\psi\right)
\label{S^min_fr}
\end{align}
with the constraints (\ref{constraint for pure YM on G/H}) and
\begin{align}
[\hat{L}_a,\psi]=\frac{i}{4}f_{a\alpha\beta}\gamma^{\alpha\beta}\psi,\;\;\;
[\hat{L}_a,\bar{\psi}]=-\frac{i}{4}f_{a\alpha\beta}\bar{\psi}\gamma^{\alpha\beta}.
\label{constraint for psi 4} 
\end{align} 
We expand the reduced model around a classical solution $\hat{X}_{\alpha}=\hat{L}_{\alpha}$ and
take the limit (\ref{limit for reduced model}) to retrieve the planar limit of one-flavor fermion on $G/H$.
The reduced model
respects the right $G$ symmetry and the gauge symmetry of one-flavor fermion on $G/H$.

Finally, let us see that the theory (\ref{S_f'}) with the constraints (\ref{constraint for X_alpha}) and 
(\ref{constraint for psi 3}) is indeed one-flavor theory on $G/H$.
It is easy to verify that (\ref{S_f'}) is invariant under the transformation
\begin{align}
&X_{\alpha}\rightarrow X_{\alpha}+\epsilon({\cal L}_aX_{\alpha}-if_{a\alpha\beta}X_{\beta}), \\
&\psi\rightarrow \psi+\epsilon\left({\cal L}_a\psi-\frac{i}{4}f_{a\alpha\beta}\gamma^{\alpha\beta}\psi\right),\;\;\;
\bar{\psi}\rightarrow 
\bar{\psi}+\epsilon\left({\cal L}_a\bar{\psi}+\frac{i}{4}f_{a\alpha\beta}\bar{\psi}\gamma^{\alpha\beta}\right).
\label{transformation for minimal fermion} 
\end{align}
We can, therefore, impose (\ref{constraint for X_alpha}) and (\ref{constraint for psi 3})
on (\ref{S_f'}) to truncate (\ref{S_f'}) consistently to a theory on  $G/H$.
We will check below that the resulting theory is the one with one-flavor Dirac fermion on $G/H$.
We define $\tilde{\rho}(t_a)$ by
\begin{align}
\tilde{\rho}(t_a)=-\frac{i}{4}f_{a\alpha\beta}\gamma^{\alpha\beta}.
\end{align}
$\tilde{\rho}(t_a)$ satisfies
\begin{align}
&[\tilde{\rho}(t_a),\tilde{\rho}(t_b)]=if_{abc}\tilde{\rho}(t_c), \\
&e^{-i\theta^a(y)\tilde{\rho}(t_a)}\gamma_{\alpha}e^{i\theta^b(y)\tilde{\rho}(t_b)}=Ad(h)_{\alpha\beta}\gamma_{\beta}.
\end{align}
We can solve (\ref{constraint for psi 3}) as 
\begin{align}
\psi=e^{i\theta^a(y)\tilde{\rho}(t_a)}\chi(\sigma), \;\;\;\bar{\psi}=\bar{\chi}(\sigma)e^{-i\theta^a(y)\tilde{\rho}(t_a)}.
\label{solution of constraint for minimal fermion}
\end{align}
Substituting (\ref{solution of constraint for X_alpha}) and 
(\ref{solution of constraint for minimal fermion}) into (\ref{S_f'}) indeed yields
\begin{align}
S^{min}_f=-\frac{w}{\kappa^2}\int d^{D-d}\sigma\sqrt{K}\:\mbox{Tr}
\left(e^{\mu}_{\alpha}\bar{\chi}\gamma^{\alpha}\left(\partial_{\mu}\chi
+\frac{1}{4}\omega^{\beta\gamma}_{\mu}\gamma_{\beta\gamma}\chi+i[a_{\mu},\chi]\right)+m_f\bar{\chi}\chi\right).
\label{S^min_f}
\end{align}

\section{CS-like theories on $G$ and $G/H$}
In this section, we construct CS-like theories on $G$ and $G/H$ and give their reduced models.
The CS 3-form on $G$ is defined by
\begin{align}
\omega_3=\mbox{Tr}\left(A\wedge dA+\frac{2i}{3}A\wedge A\wedge A\right).
\end{align}
For an arbitrary $N\times N$ unitary matrix, the gauge transformation is given by
\begin{align}
A'=idUU^{-1}+UAU^{-1}.
\label{gauge transformation}
\end{align}
As is well known, the CS 3-form is transformed under the gauge transformation as
\begin{align}
\omega_3'=\omega_3-id\mbox{Tr}(U^{-1}dU\wedge A)
-\frac{1}{3}\mbox{Tr}(dUU^{-1}\wedge dUU^{-1}\wedge dUU^{-1}).
\label{gauge transformation of omega_3}
\end{align}
The 3-form in the third term of RHS is closed:
\begin{align}
d\mbox{Tr}(dUU^{-1}\wedge dUU^{-1}\wedge dUU^{-1})=0,
\label{closed 3-form}
\end{align}
which means that the 3-form belongs to $H^3(G)$.

We define a 3-form $f$ on $G$ in terms of the structure constant $f_{ABC}$:
\begin{align}
f=\frac{1}{3!}f_{ABC}E^A\wedge E^B\wedge E^C.
\label{f}
\end{align}
It is easy to show that
\begin{align}
&df=0, \label{closed f}\\
&d\ast f=0,
\label{closed *f}
\end{align}
which means that $f$ and $\ast f$ are harmonic forms so that $f$ and $\tilde{f}$ are non-zero elements
of $H^3(G)$ and $H^{D-3}(G)$, respectively.
We define the CS-like theory on $G$
\begin{align}
S=\frac{1}{\alpha}\int \omega_3\wedge \ast f.
\label{Chern-Simons-like theory on G}
\end{align}
We can show that (\ref{Chern-Simons-like theory on G}) has the gauge symmetry as follows. 
Using (\ref{gauge transformation of omega_3}), (\ref{closed 3-form}),
(\ref{closed *f}) and the Poincare duality, we find that $S$ transforms to
\begin{align}
S'=S-\frac{1}{3\alpha}\int_{C_3}\mbox{Tr}(dUU^{-1}\wedge dUU^{-1}\wedge dUU^{-1}),
\end{align}
where $C_3$ is the 3-cycle dual to $*f$. As in the case of 
three-dimensional CS theory, if we normalize $\alpha$ appropriately, we obtain
\begin{align}
S'=S+2\pi n
\end{align}
for an integer $n$, so that $e^{iS}$ is indeed invariant.

(\ref{Chern-Simons-like theory on G}) is rewritten as
\begin{align}
S&=\frac{1}{6\alpha}\int d^Dx \sqrt{G}\: E^M_AE^N_BE^L_C f^{ABC}
\mbox{Tr}\left(A_M\partial_NA_L+\frac{2i}{3}A_MA_NA_L\right) \nonumber\\
&=\frac{1}{6\alpha}\int dg \: f^{ABC}\mbox{Tr}\left(iX_A{\cal L}_BX_C+\frac{1}{2}f_{BCD}X_AX_D
+\frac{2i}{3}X_AX_BX_C\right).
\label{Chern-Simons-like theory on G 2}
\end{align}
The reduced model of (\ref{Chern-Simons-like theory on G 2}) is
\begin{align}
S_r=\frac{v}{6\alpha}f^{ABC}\mbox{Tr}\left(i\hat{X}_A[\hat{L}_B,\hat{X}_C]+\frac{1}{2}f_{BCD}\hat{X}_A\hat{X}_D
+\frac{2i}{3}\hat{X}_A\hat{X}_B\hat{X}_C\right),
\label{reduced model of Chern-Simons-like theory}
\end{align}
which retrieves the planar limit of (\ref{Chern-Simons-like theory on G}) in the limit (\ref{limit for reduced model}).
By making the redefinition $\hat{L}_A+\hat{X}_A\rightarrow \hat{X}_A$, we obtain from
(\ref{reduced model of Chern-Simons-like theory}) up to an irrelevant constant term
\begin{align}
S_r'=\frac{v}{6\alpha}f^{ABC}\mbox{Tr}\left(\frac{1}{2}f_{BCD}\hat{X}_A\hat{X}_D
+\frac{2i}{3}\hat{X}_A\hat{X}_B\hat{X}_C\right).
\label{reduced model of Chern-Simons-like theory 2}
\end{align}
$\hat{X}_A=\hat{L}_A$ is a classical solution of (\ref{reduced model of Chern-Simons-like theory 2}).
We expand (\ref{reduced model of Chern-Simons-like theory 2}) around $\hat{X}_A=\hat{L}_A$ and
take the limit (\ref{limit for reduced model}). Then, (\ref{reduced model of Chern-Simons-like theory 2})
retrieves the planar limit of the original CS-like theory. 
For $G\simeq SU(2)$, (\ref{Chern-Simons-like theory on G}) is nothing but pure CS theory on the 3-sphere\footnote{
In \cite{Ishiki:2009vr,Ishiki:2010pe}, the different type of the large $N$ reduction on $S^3$
developed in \cite{Ishii:2008ib} was explicitly demonstrated for this theory.}.

Next, we study the CS-like theory on $G/H$. It is easy to see from (\ref{Lie derivative of E^A_M}) that
\begin{align}
\delta_A(f^{BCD}E^M_BE^N_CE^L_D)=0.
\end{align}
This implies that (\ref{Chern-Simons-like theory on G 2}) is invariant under the transformation
(\ref{transformation of A_M}). 
Hence, by imposing the constraints (\ref{X_a=0}) and (\ref{constraint for X_alpha}) 
on (\ref{Chern-Simons-like theory on G 2}), we can truncate
(\ref{Chern-Simons-like theory on G 2}) to a theory on $G/H$ as in the case of YM theory on $G$.
The resulting theory is a CS-like theory on $G/H$ which takes the form
\begin{align}
S^{G/H}&=\frac{w}{6\alpha}\int d^{D-d}\sigma\sqrt{K} \: f^{\alpha\beta\gamma}\mbox{Tr}
\left(iX_{\alpha}{\cal L}_{\beta}X_{\gamma}+\frac{1}{2}f_{\beta\gamma\delta}X_{\alpha}X_{\delta}
+\frac{2i}{3}X_{\alpha}X_{\beta}X_{\gamma}\right) \nonumber\\
&=\frac{w}{\alpha}\int \tilde{w}_3\wedge \ast \tilde{f}.
\label{Chern-Simons-like theory on G/H}
\end{align}
Here $\tilde{\omega}_3$ is the CS 3-form on $G/H$:
\begin{align}
\tilde{\omega}_3=\mbox{Tr}\left(a\wedge \tilde{d}a+\frac{2i}{3}a\wedge a\wedge a\right)
\end{align}
with
\begin{align}
&a=a_{\mu}d\sigma^{\mu},\nonumber\\
&\tilde{d}=d\sigma^{\mu}\frac{\partial}{\partial \sigma^{\mu}}.
\end{align}
$\tilde{f}$ is a 3-form on $G/H$, which is analogous to $f$ on $G$:
\begin{align}
\tilde{f}=f_{\alpha\beta\gamma}e^{\alpha}\wedge e^{\beta}\wedge e^{\gamma}.
\end{align}
$\ast$ stands for the Hodge dual on $G/H$. (\ref{solution of constraint for X_alpha}) has been used
to obtain the second line of (\ref{Chern-Simons-like theory on G/H}).
By construction, (\ref{Chern-Simons-like theory on G/H}) has the symmetry under the gauge transformation 
(\ref{gauge transformation}) with $\delta_aU=0\;\;(\partial U/\partial y^m=0)$. 
Indeed, we can easily show that
\begin{align}
\tilde{d}\ast \tilde{f}=0
\end{align}
which means that $\ast \tilde{f}\in H^{D-d-3}(G/H)$.
Hence, under the gauge transformation
\begin{align}
a'=i\tilde{d}uu^{-1}+uau^{-1}
\end{align}
with $u$ an arbitrary $\sigma$-dependent $N\times N$ unitary matrix, (\ref{Chern-Simons-like theory on G/H})
is transformed as $S'=S+2\pi n$, as in the case of the CS-like theory on $G$.

The reduced model of (\ref{Chern-Simons-like theory on G/H}) is
\begin{align}
S^{G/H}_r=\frac{v}{6\alpha}f^{\alpha\beta\gamma}\mbox{Tr}\left(i\hat{X}_{\alpha}[\hat{L}_{\beta},\hat{X}_{\gamma}]
+\frac{1}{2}f_{\beta\gamma\delta}\hat{X}_{\alpha}\hat{X}_{\delta}+\frac{2i}{3}\hat{X}_{\alpha}\hat{X}_{\beta}\hat{X}_{\gamma}\right).
\label{S^G/H_r}
\end{align}
with the constraint (\ref{constraint for pure YM on G/H}). The redefinition $\hat{L}_{\alpha}+\hat{X}_{\alpha}\rightarrow \hat{X}_{\alpha}$
in (\ref{S^G/H_r}) leads to
\begin{align}
S^{G/H}_r{}'=\frac{v}{6\alpha}f^{\alpha\beta\gamma}\mbox{Tr}\left(
\frac{1}{2}f_{\beta\gamma\delta}\hat{X}_{\alpha}\hat{X}_{\delta}+\frac{2i}{3}\hat{X}_{\alpha}\hat{X}_{\beta}\hat{X}_{\gamma}
+f_{\beta\gamma a}\hat{L}_a\hat{X}_{\alpha}\right)
\end{align}
up to an irrelevant constant term.

\section{Conclusion and discussion}
In this paper, we showed that the large $N$ reduction holds on coset spaces. The reduced models of large $N$ field theories
on coset spaces are obtained by imposing the constraints on the reduced models of the corresponding theories on group
manifolds. 
We also constructed CS-like theories on group manifolds and coset spaces, and gave their reduced models.

As an application of our findings in this paper, we can define large $N$ field theories 
on $S^4\simeq SO(5)/SO(4)$ non-perturbatively in terms of their reduced models.  
In particular, it is interesting
to construct the reduced models of supersymmetric gauge theories on $S^4$.
While the reduced models of those on $R\times S^3$ constructed in \cite{Ishii:2008ib,KST} still
has the continuous time direction, the reduced models of those on $S^4$ are indeed defined in zero dimension
so that they would be more tractable. The large $N$ reduction for CS-like theories can be applied to the study
of the ABJM theory \cite{Aharony:2008ug}.

We hope to find reduced models of large $N$ field theories on a wider class of curved spaces and eventually to make 
progress in the description of curved space-times in the matrix models conjectured to give a non-perturbative formulation of superstring.

\section*{Acknowledgment}
This work was supported by the Grant-in-Aid for the Global COE program ``The Next Generation
of Physics, Spun from Universality and Emergence" from the Ministry of Education, Culture,
Sports, Science and Technology (MEXT) of Japan.
The work of S.~S. 
is supported 
by JSPS.
The work of A.~T.
is supported 
by the Grant-in-Aid for Scientific Research (19540294) from JSPS.

\end{document}